\documentclass[a4paper,11pt]{article}
\pdfoutput=1 

\usepackage{jinstpub} 

\title{\boldmath The FAIR Phase 0 program of the CBM TOF}

\author[a,1]{I. Deppner,\note{Corresponding author.}}
\author[a]{N. Herrmann,}

\affiliation[a]{Physikalisches Institut der Universit\"at Heidelberg, Im Neuenheimer Feld 226, Heidelberg, Germany}

\emailAdd{deppner@physi.uni-heidelberg.de}

\abstract{In order to provide particle identification (PID) of charged hadrons at the future high-rate Compressed Baryonic Matter (CBM) experiment, the TOF group has developed a large-area Time-of-Flight (ToF) wall equipped with Multi-gap Resistive Plate Chambers (MRPC) capable to operate at incident particle fluxes of up to 30 kHz/cm$^2$. Prior to its destined operation at the Facility for Antiproton and Ion Research (FAIR) - starting in 2025 - this high-rate timing MRPC technology will be used for physics research at two scientific pillars of the FAIR Phase-0 program: the end-cap TOF upgrade of the STAR experiment at RHIC and the mini-TOF (mTOF) wall of the mini-CBM (mCBM) experiment at Schwerionensynchrotron18 (SIS18). At STAR, the fixed-target program of the Beam Energy Scan II (BES-II) will rely on 108 CBM MRPC detectors for forward PID at trigger rates of up to 2 kHz. At mCBM, high-performance benchmark runs of Lambda-baryon production at top SIS18 energies and CBM design interaction rates of 10 MHz will become feasible with a PID backbone consisting of 25 CBM MRPC detectors. Apart from the physics perspective, these pre-FAIR involvements will help gathering experience in operating the final CBM TOF wall comprising about 1500 MRPC detectors and 110,000 readout channels. The status of the FAIR phase 0 program will be discussed.}


\arxivnumber{} 

\collaboration[c]{on behalf of the CBM collaboration}

\proceeding{XV Workshop on Resistive Plate Chambers and related detectors$^{\text{th}}$ \\
	10 - 14 February, 2020,\\
	Tor Vergata, Rome, Italy}

\usepackage{hyperref}

\begin{document}
	\maketitle
	\flushbottom
	
	\section{Introduction}
	\label{sec:intro}
	
	The Compressed Baryonic Matter spectrometer (CBM) is a future fixed-target heavy-ion experiment being build at the Facility for Anti-proton and Ion Research (FAIR) in Darmstadt, Germany  \cite{CBM_webpage}. The CBM collaboration aims to explore the phase diagram of strongly interacting matter at high baryon densities in the beam energies range from 2 A GeV to 11 A GeV with Au + Au collisions at unprecedented interaction rates of up to 10~MHz~\cite{Ablyazimov:2017guv}. This requires very fast and radiation hard detectors, a novel data read-out and analysis concept including free streaming front-end electronics, and a high performance computing cluster for online event selection \cite{Senger2018}. In order to create an environment where all aspects of the developed detector systems can be elaborated under real battle conditions the FAIR Phase 0 program was launched. The idea is to install and operate existing FAIR related detector equipment in running experiments all over the world.
	
	CBM TOF subsystem participates in the following two FAIR Phase 0 programs:
	
	\begin{itemize}
		\item mTOF at mCBM/SIS18, comprising 1600 readout channels, to test high rate characteristics and system integration aspects
		\item eTOF at STAR/BNL, comprising 6912 readout channels, to test long term stability and produce physics results
	\end{itemize}

	In the following the status of these two FAIR phase 0 programs will be discussed.
	
\begin{figure}[htbp]
		\centering 
		\includegraphics[width=.47\textwidth,origin=c]{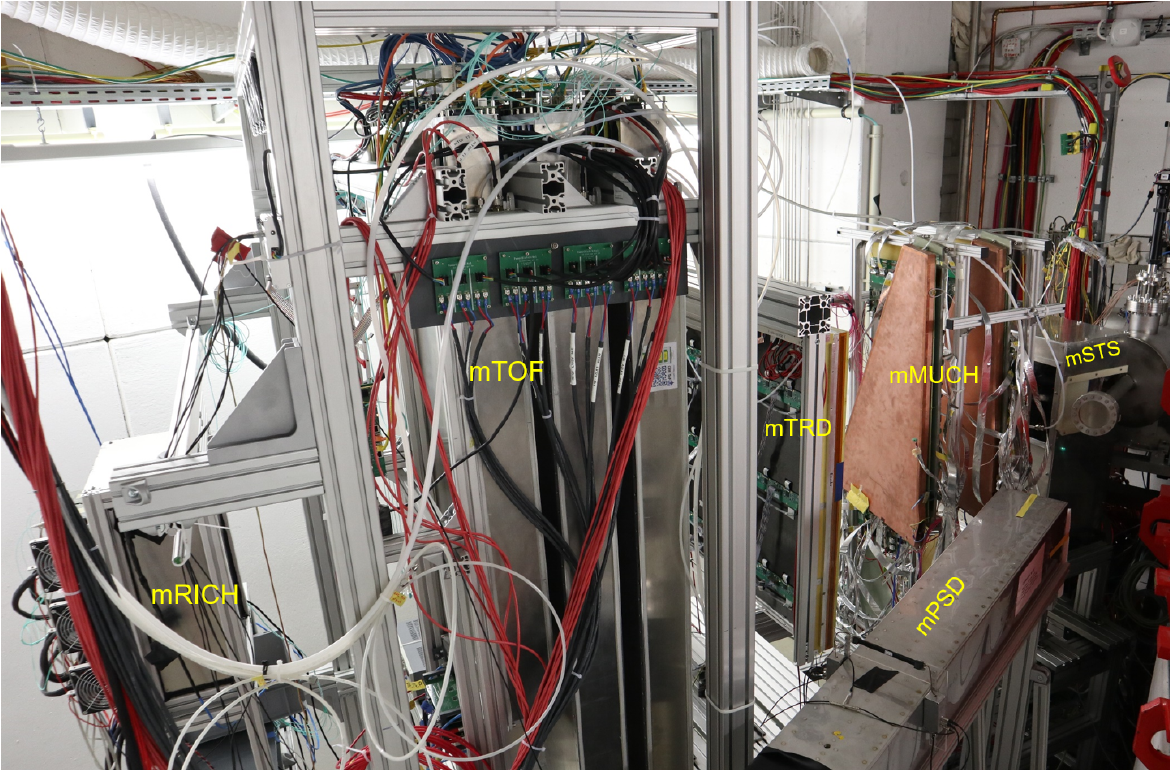}
		\includegraphics[width=.45\textwidth,origin=c]{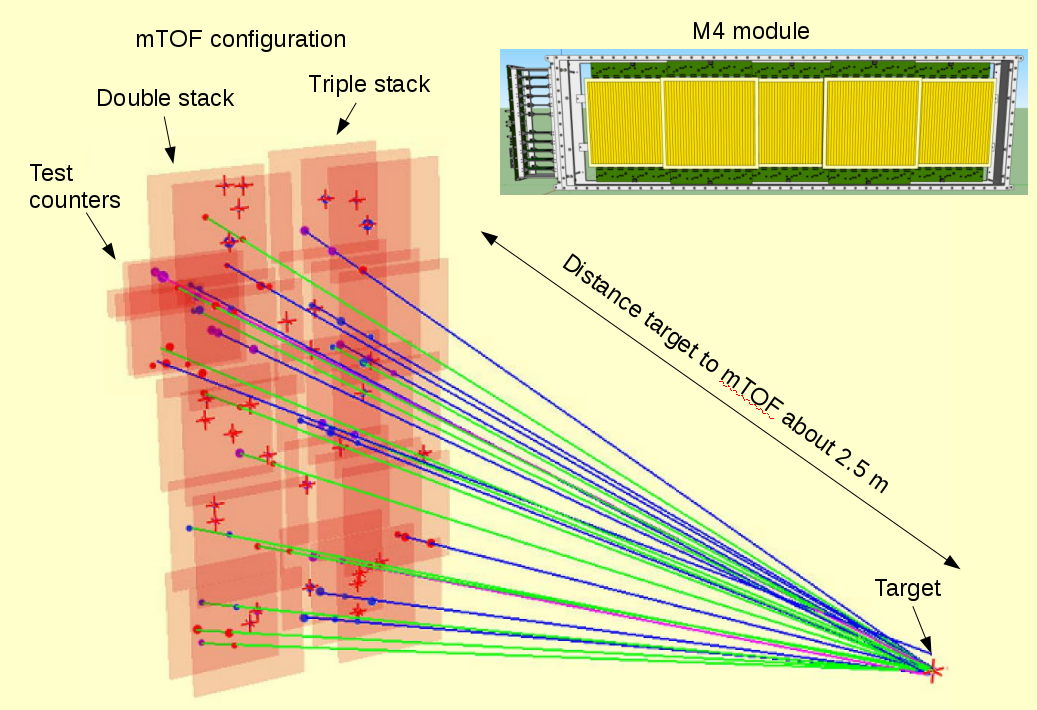}
		
		\caption{\label{fig:1} Left: mCBM setup at SIS18/GSI. In the center mTOF is visible together with other subsystems. Right: Event display of a high multiplicity Pb on Au event generating 100 hits in mTOF of which 28 tracks. The M4 modules are arranged in a triple and a double stack configuration.}
	\end{figure}

	\section{mTOF at mCBM}
	\label{sec:mTOF}
	
	mCBM is a full-system test-setup for CBM located at the SIS18 synchrotron at GSI/Germany. The primary aim is to study, commission and test the complex interplay of the different detector systems with the free-streaming data acquisition and the fast online event reconstruction and selection. It will allow to test the detector and electronics components developed for the CBM experiment as well as the corresponding online/offline software packages under realistic experiment conditions up to top CBM interaction rates of 10 MHz \cite{mCBM_webpage,mCBM_Beamtime_appl}. The left side in Fig.~\ref{fig:1} depicts the mCBM setup with all subsystems. In the center the mini-Time-of-Flight wall (mTOF) is illustrated. It consists out of 5 so called M4 CBM full size modules (see right side of Fig.~\ref{fig:1}), housing 5 MRPC2 (see Table~\ref{Table}, formerly called MRPC3a \cite{Deppner2019,Lyu2019}), counter each. Each module has an active area of 152$\times$27 cm$^2$ and comprises 320 read out channels. As depicted in Fig.~\ref{fig:1} the 5 mTOF modules were arranged in a triple and a double stack configuration. The triple stack, centered at 25$^{\circ}$ in respect to the beam, allows to form triple hit tracklets, which can be used by the other subsystem as a reference. However, with three staggered modules the performance of each module can be elaborated by using the other two as a reference (see contribution \cite{Zhang2020} this proceeding). The double stack, located closer to the beam (at a polar angle of about 12$^{\circ}$), faces a higher particle flux and offers a reference system to additional CBM TOF MRPC prototypes, which were tested during the beam time campaigns. It is evident that also from the doubles stack, two hit tracklets can be obtained. The right side of Figure~\ref{fig:1} shows an event display of a high multiplicity Pb on Au event generating 100 hits in mTOF of which 28 tracks could be reconstructed, demonstrating the capability of this facility to test MRPC detectors at high rates in a multi-hit environment and under full counter illumination.
	
\begin{figure}[htbp]
		\centering 
		\includegraphics[width=.40\textwidth,origin=c]{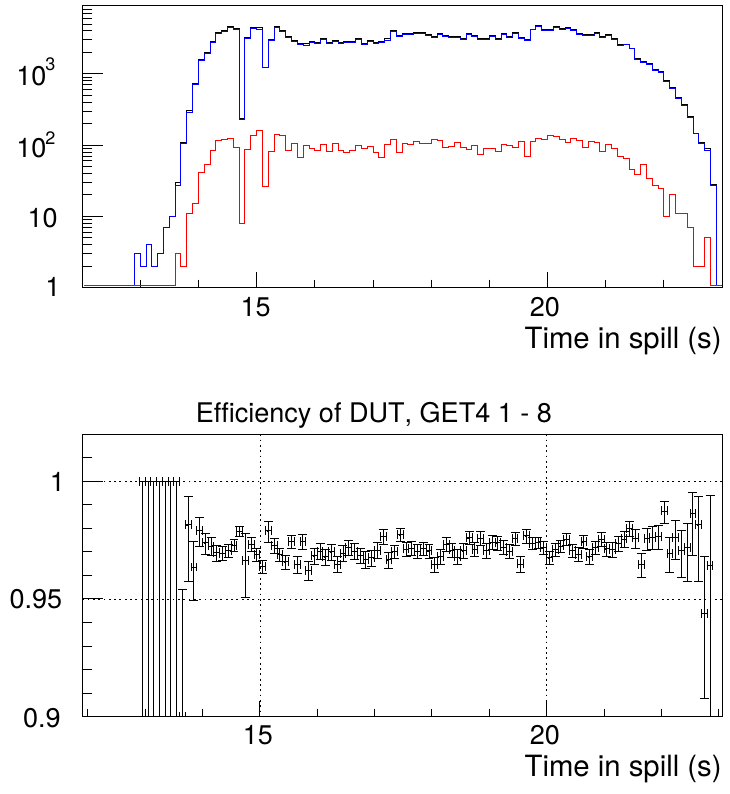}
		\includegraphics[width=.49\textwidth,origin=c]{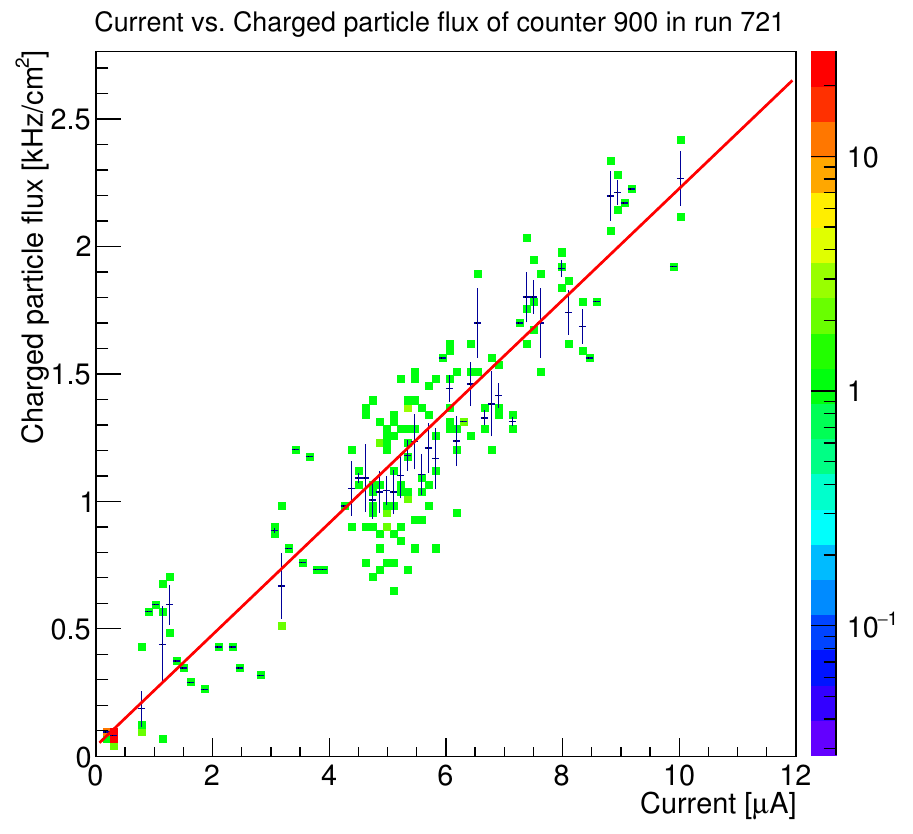}	
		\caption{\label{fig:2} Top left: found (blue) and missed (red) hits on counter 900 as function of time in spill. During run 717 the incoming particle flux was in the order of 600 Hz/cm$^2$.	Bottom left: Efficiency as function of time in spill. Right: Correlation between incoming charged particle flux and drawn current of counter 900 (active area 864 cm$^2$). In run 721 particle fluxes up to 2.3 kHz/cm$^2$ were reached.}
	\end{figure}

	During the beam time campaign in March 2020 a chamber housing two MRPC3 prototypes (see Table~\ref{Table}, former MRPC3b \cite{Deppner2019,Hu2019}), equipped with ultra thin float glass electrodes (here called counter 900 and 901), was installed behind the double stack. The aim was to measure the rate capability of these counters which are supposed to be located at the outermost region of the CBM TOF wall \cite{Deppner2012} where the incident particle flux is below 1 kHz/cm$^2$. More information about the MRPC2/3 prototypes is given in Table~\ref{Table}. With the tracking method described in \cite{Zhang2020} the efficiency of counter 901 was evaluated for a run where the incident particle flux was about 600 Hz/cm$^2$. Figure~\ref{fig:2} upper left shows the number of found hits (blue) and the number of missed hits (red) as function of time in spill (bin size 0.1~s). By dividing the missed hits by the sum of the missed and found hits the efficiency for each bin was obtained (shown in the lower left Fig.~\ref{fig:2}). It results in about 97~\% without a visible degradation with time in spill. Analysis of runs taken at higher interaction rates is ongoing. As a precursor, the right side of Fig.~\ref{fig:2} shows the  normalized hit count rate of counter 900 plotted versus the drawn HV current. A linear dependence is observed up to an incident particle flux of about 2.3 kHz/cm$^2$ which could be an indication that the counter is functional up to this rate. Based on the measured current of all involved counters, a maximal rate of about 8 kHz/cm$^2$ during this beam time could be estimated, which is sufficient to evaluate even the MRPC2 counters with low resistivity glass electrodes.
		\begin{table}	        
			\begin{center}
				\begin{tabular}{|| c | c | c ||} 
					\hline
					RPC type & MRPC2 &  MRPC3 \\
					\hline\hline
					Active area & 32 $\times$ 27 cm$^2$ & 32 $\times$ 27 cm$^2$ \\ 
					\hline
					Nr. of strips / pitch & 32 / 1 cm & 32 / 1 cm \\
					\hline
					Nr. of gaps & 2 $\times$ 4 & 2 $\times$ 5 \\
					\hline
					Gap size & 250 $\mu$m & 230 $\mu$m \\
					\hline
					Glass type & low resistivity & float \\
					\hline
					Glass thickness & 700 $\mu$m & 280 $\mu$m \\
					\hline	
				\end{tabular}
			\end{center}     
			\caption{Key parameters of the MRPC2 and MRPC3 prototype.}
			\label{Table}
		\end{table}
		
	\section{eTOF at STAR}
	\label{sec:eTOF}
	
\begin{figure}[htbp]
		\centering 
		\includegraphics[width=.60\textwidth,origin=c,angle=270]{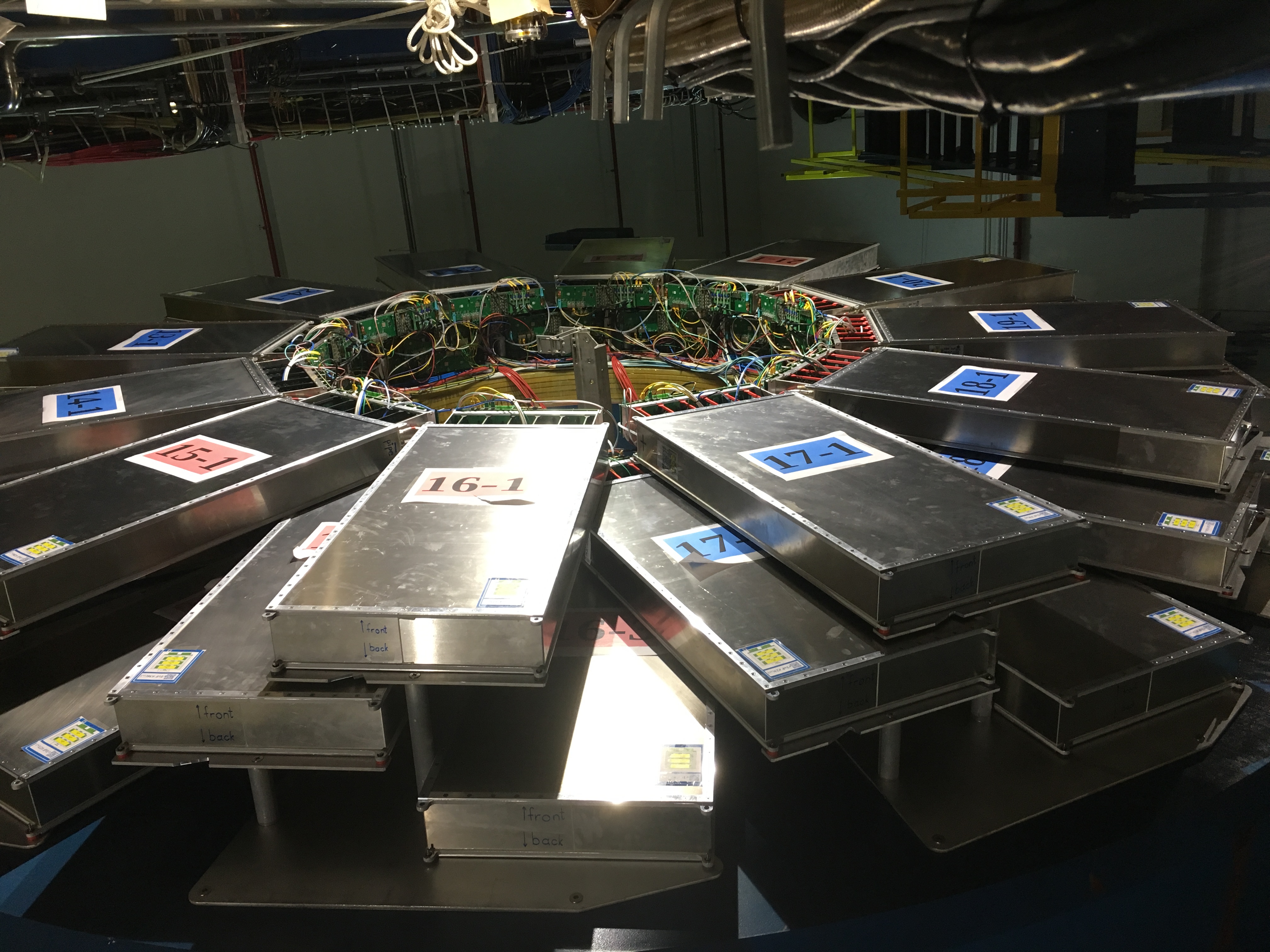}
		
		\caption{\label{fig:3} Photograph of the eTOF wheel.}
	\end{figure}

	The endcap Time-of-Flight (eTOF) project comprises the installation, commissioning and operation of CBM TOF modules positioned at the east magnet pole tip of the STAR apparatus (see Fig.~\ref{fig:3}) during the Beam Energy Scan II (BESII) campaign and the participation in the analysis of data obtained in runs with eTOF. The eTOF upgrade offers a PID extension in pseudorapidity between $-1.6<\eta<-1.1$ for the collider collision mode. For the fix target collision mode, the PID extension of the pseudorapidity, which is between $1.6<\eta<2.1$, is absolutely essential in order to cover the mid-rapidity range (cf. Table~\ref{Table1}). Note that the sign of the pseudorapidity is at STAR by definition negative in east direction, however, for the fix target mode it is redefined to positive values. 
	
	eTOF consists of 36 modules grouped in 12 sectors which are arranged in 3 layers in a spokes structure inside a wheel around the beam axis. 12 modules are equipped with counters of type MRPC2 while 24 modules have counters of type MRPC3. Each module houses 3 MRPCs which leads to a total of 108 counters and 6912 read out channels. The full hardware installation was completed in Nov. 2018. After a commissioning phase of about 10 weeks the first data taking started in Feb. 2019 by recording about 580 M Au+Au events at $\sqrt{s_{NN}}=11.5\,GeV$ with an eTOF participation of 85\%.
	
	Table~\ref{Table1} summarizes the data sets taken during the STAR Beam Energy Scan II campaign running from December 2019 to March 2020. The 9.2 GeV collider run could not be finished due to the Corona shut down. All fix target runs were successfully completed and about 100~M events with eTOF data were collected each.
	
	\begin{table}	        
		\begin{center}
			\begin{tabular}{|| c | c | c | c | c | c ||} 
				\hline
				Mode & $\sqrt{s_{NN}}$ & $E_{beam}$  & mid-rapidity  & \# events & Status \\
				&  [GeV] & [GeV]  & $y_{CM}$ &with eTOF & \\  
				\hline\hline
				Col. & 11.5 & 5.75 & 0 &127 M & finished \\ 
				\hline
				Col. & 9.2 & 4.6 & 0 &\~ 30 M & not finished \\
				\hline
				FXT & 7.7 & 31.2 & 2.03 &100 M & finished \\
				\hline
				FXT & 6.2 & 19.5 & 1.86 &80 M & finished \\
				\hline
				FXT & 5.2 & 13.5 & 1.68 &89 M & finished \\
				\hline
				FXT & 4.5 & 9.8 & 1.52 &106 M & finished \\
				\hline
				FXT & 3.9 & 7.3 & 1.37 &106 M & finished \\
				\hline
				FXT & 3.5 & 5.75 & 1.25 &100 M & finished \\ 
				\hline	
			\end{tabular}
		\end{center}     
		\caption{Data sets taken during the STAR Beam Energy Scan II campaign running from December 2019 to March 2020 }
		\label{Table1}
	\end{table}
	
\begin{figure}[tb]
		\centering
		\includegraphics[width=.47\textwidth,origin=c]{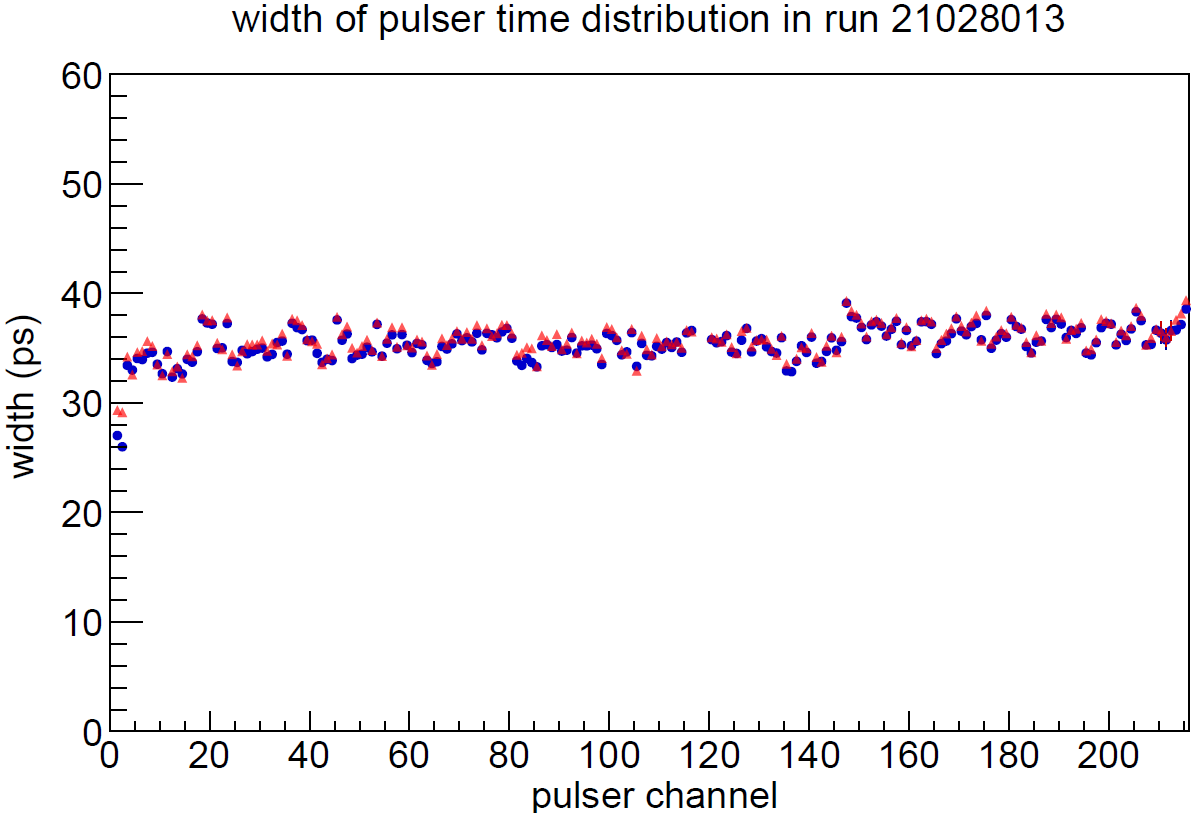}
		\includegraphics[width=.50\textwidth,origin=c]{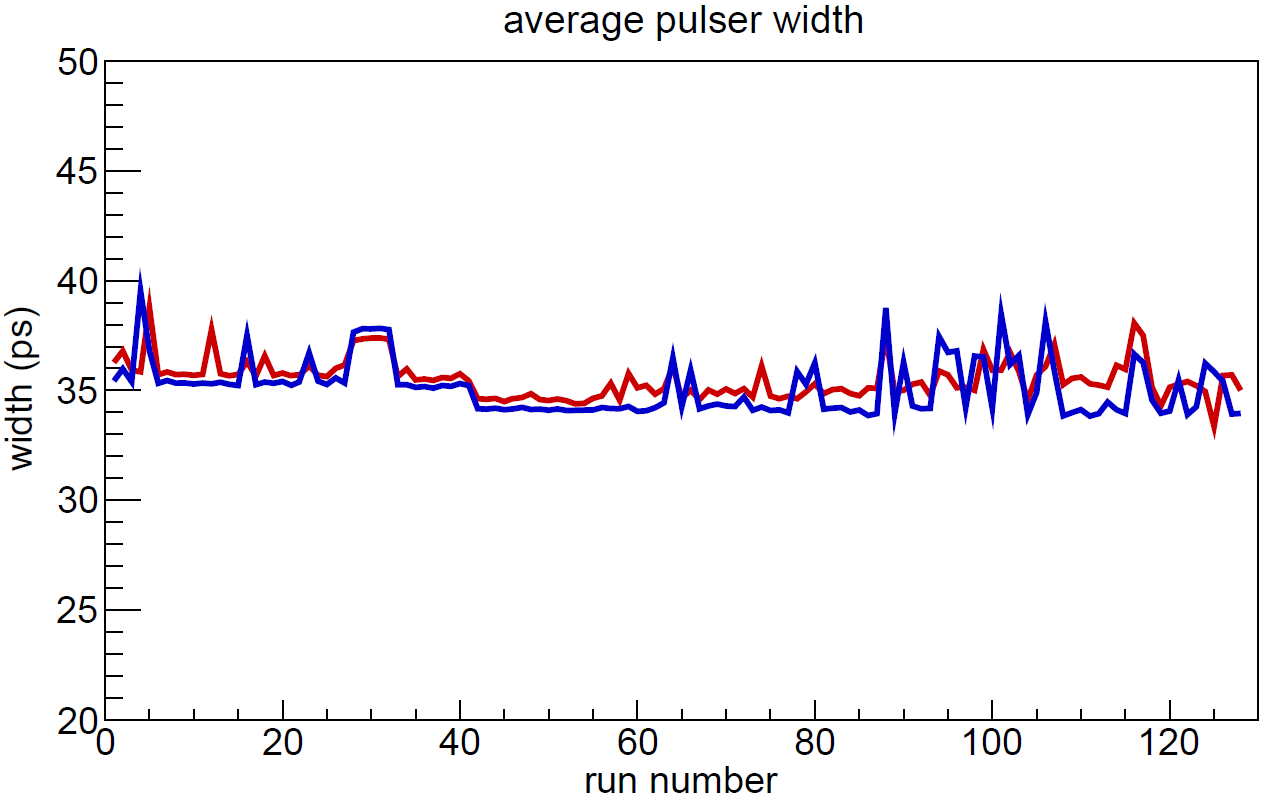}
		\caption{\label{fig:4}Left: width of the time distribution obtained by measuring the arrival time of injected pulser signals on every GET4 board. Right: mean of the time distribution width vs. the run number. Figures from \cite{Seck2020}.} 
	\end{figure}

\begin{figure}[tb]
		\centering
		\includegraphics[width=.90\textwidth,origin=c]{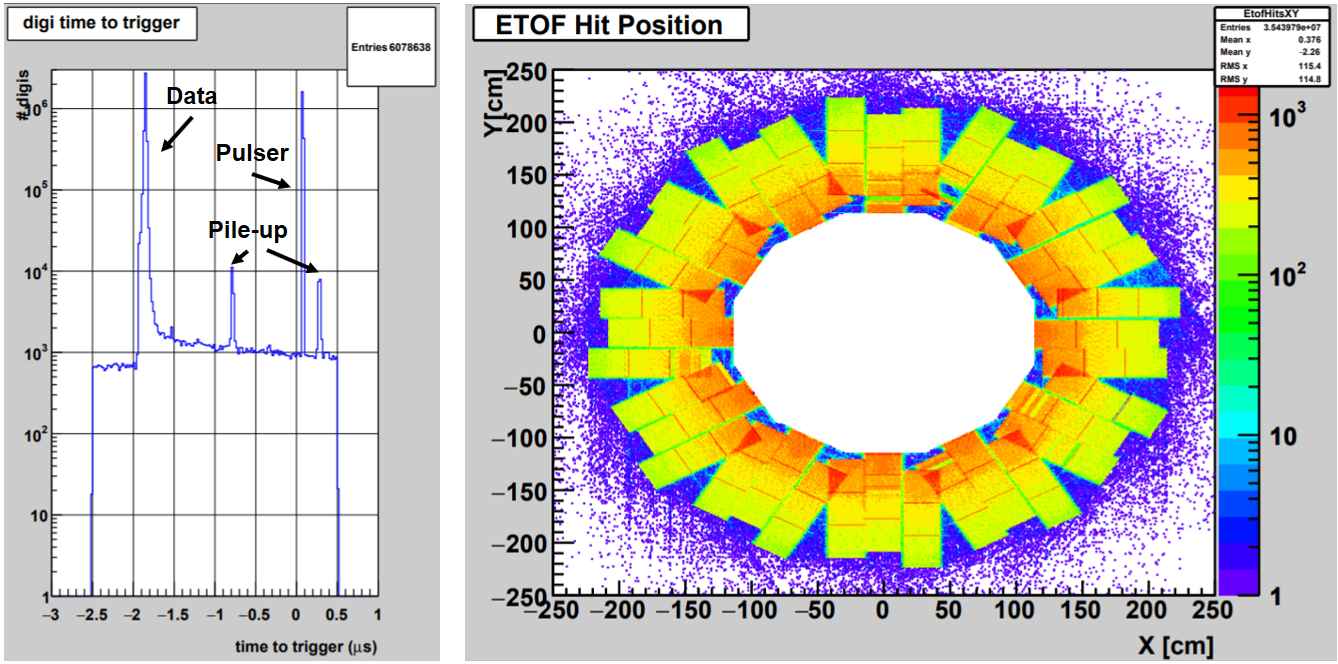}	
		\caption{\label{fig:5}Left: Data interval of $\mu$s sent to the STAR event builder. The main peak corresponds to the data. Right:  eTOF hit distribution monitored by the online QA plot obtained for one $\sqrt{s_{NN}}=7.7\,GeV$ fix target run.} 
	\end{figure}

	The readout system for eTOF is using CBM's free-streaming architecture and CBM's hardware and software components. It comprises 216 PADI (preamplifier and discriminator) and 216 GET4 (TDC) boards (for 108 counters). From each module 2$\times$20 m long duplex fibers transport the data which were accumulated by 2 GBTx chips to a $\mu$TCA crate housing 12 FPGA based data concentrator and data pre-processing boards called AFCKs (one for each sector). From here the data are sent via 12$\times$100~m long duplex fibers to the FPGA based FLes Interface Board (FLIB) located in the DAQ room of STAR. In comparison to 2019 an improved clock distribution method was installed offering a system synchronization with a jitter in the order of 35 ps over the full wheel (see left plot in Fig.~\ref{fig:4}). This plot shows the width of the time distribution (red corresponds to the Gaussian sigma and blue to the RMS) obtained by measuring the arrival time of injected pulser signals on every GET4 board. The stability of the system is demonstrated on the right plot of Fig.~\ref{fig:4}. Here the mean of time distribution width from all pulser channels is plotted vs. the run number. The range of 130 runs reflects a time period of several days. The online event building is performed with ZMQ in parallel on 12 CPUs and is based on the trigger token information (received from STAR) injected in the data stream at the AFCK level. A time window of 3 $\mu$s around the trigger time stamp is selected (conf. left side of Fig.~\ref{fig:5} and sent to the STAR event builder where the data are monitored and stored for offline analysis. The right side Fig.~\ref{fig:5} shows an online QA plot of the eTOF hit distribution in the global xy-plane obtained from one run (30 min) of $\sqrt{s_{NN}}=7.7\,GeV$ fix target data. Even with a preliminary calibration the MRPC hits show a uniform distribution with a rate gradient towards the center and only a tiny fraction is reconstructed outside the active area.
	
\begin{figure}[tb]
		\centering
		\includegraphics[width=.42\textwidth,origin=c]{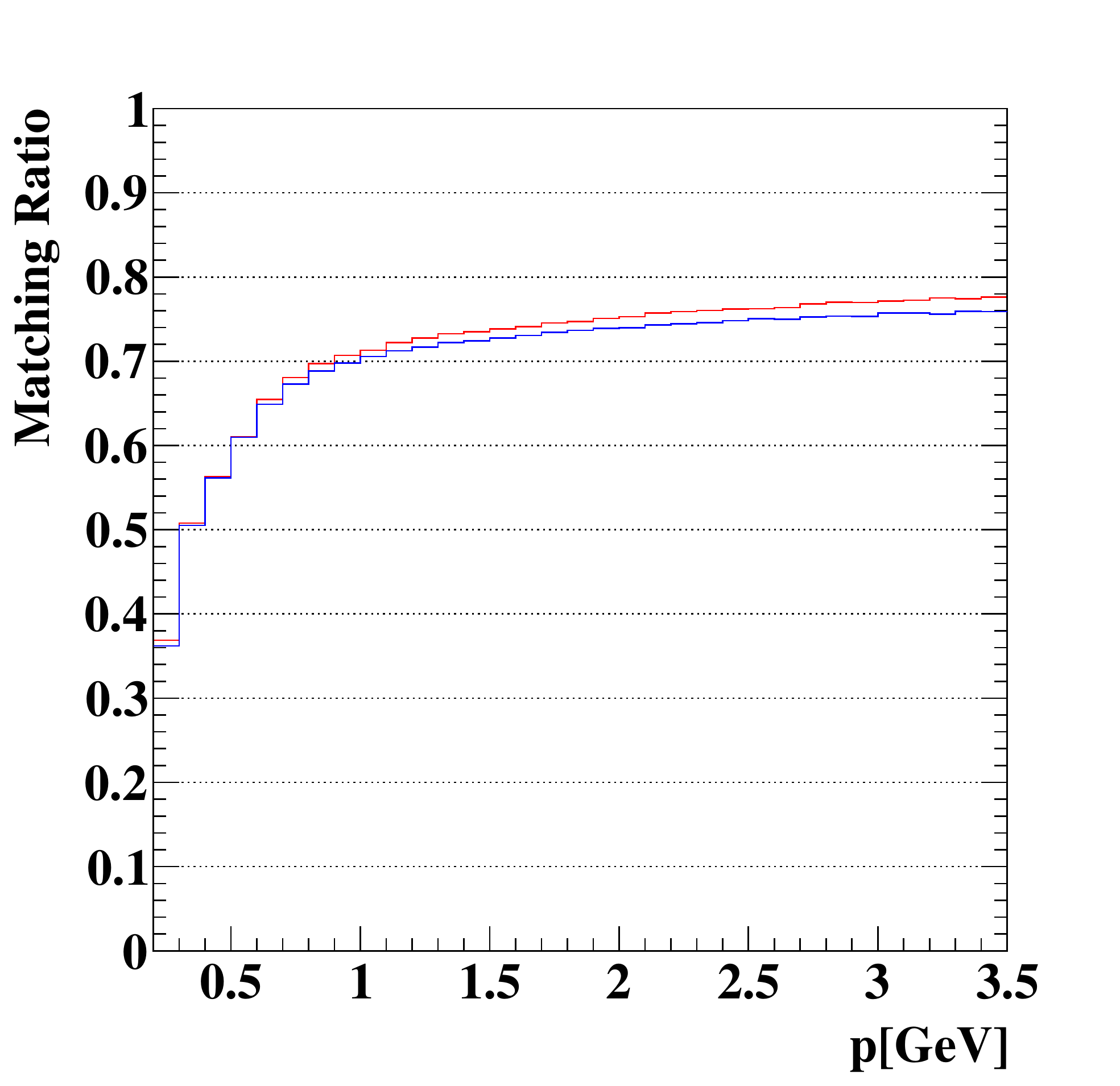}
		\includegraphics[width=.49\textwidth,origin=c]{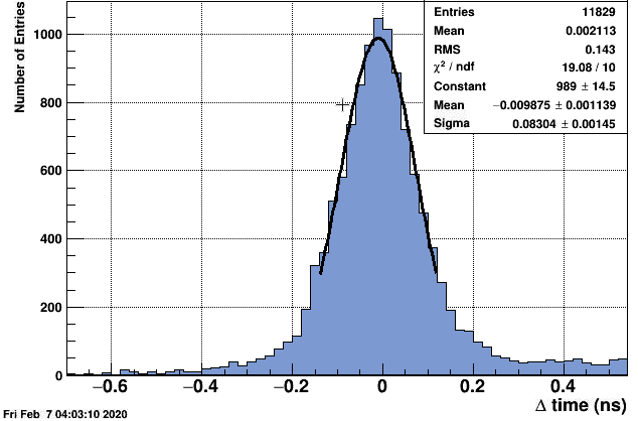}
		\caption{\label{fig:6}Left: matching efficiency of MRPC hits in respect to the extrapolated TPC tracks as function of the particle momentum. Right: System time resolution obtained with relativistic pions. Figures from \cite{Weidenkaff2020}.} 
	\end{figure}

\begin{figure}[tb]
		\centering
		\includegraphics[width=.60\textwidth,origin=c]{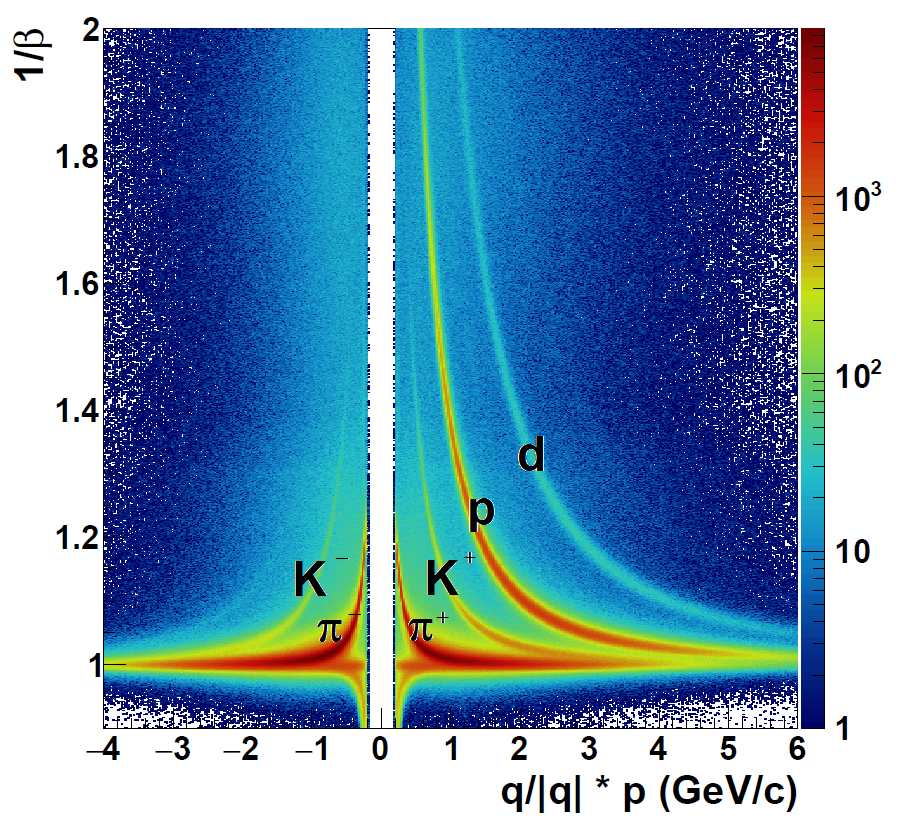}
		\caption{\label{fig:7} $1/\beta$ as funtion of particle momentum. The separation of kaons from pions up to a momenta of 2.5 GeV/c demonstrates the PID capability of eTOF. Figures from \cite{Seck2020}.} 
	\end{figure}

	After offline calibration of the TPC and eTOF the matching efficiency of MRPC hits in respect to the extrapolated TPC tracks can be deduced as function of the particle momentum (see Fig.\ref{fig:6}). The Matching radius is set to 5$\sigma$ of the measured residual width. At a momentum of 1 GeV/c a matching efficiency of 70\% is obtained. Beyond the curve levels off at 75\%. Both counter types show a similar behavior. The same picture is obtained by evaluating the system time resolution with relativistic pions. Both counter types show results in the order of 85 ps. The tail results from a background of mis-identified pions. The good time resolution is reflected in the $1/\beta$ versus the particle momenta plot shown in Fig.~\ref{fig:7} which is generated with $\sqrt{s_{NN}}=7.7\,GeV$ fix target data. The narrow particle bands allow for a kaon to pion separation of up to a momentum of 2.5~GeV/c which demonstrates the excellent PID capability of eTOF. 
	
	\section{Conclusion}
	\label{sec:conclusion}
	
	The Compressed Baryonic Matter experiment at FAIR is planned to be operational in 2025. In order to meet this target all aspects of the system have to be tested extensively. A bridge program called FAIR phase 0 has been developed where existing detector equipment is implemented and tested in running experiments. The CBM TOF project participates in two FAIR phase 0 programs: (1) mTOF in miniCBM installed at SIS/GSI and (2) eTOF in STAR/BNL. mCBM offers the opportunity to test detector equipment under conditions faced by CBM while the STAR eTOF upgrade will allow for long stability test of the full system and will produce relevant experimental data for physics analysis. Both projects are in an excellent condition and are delivering valuable output.


	
	\acknowledgments
	The authors want to express their thanks especially to the eTOF group and to the CBM and STAR collaboration. This work is partially funded through contracts 05P15VHFC1 and 05P19VHFC1 
	


\end{document}